\documentclass[preprint,12pt]{elsarticle}

\usepackage{graphicx}
\usepackage{amsmath,amssymb,amsfonts}
\usepackage{bm}
\usepackage{booktabs}
\usepackage{array}
\usepackage{tabularx}
\usepackage{float}
\usepackage[hidelinks]{hyperref}

\journal{Annals of Physics}
\biboptions{numbers,sort&compress}

\newcommand{\bea}{\begin{eqnarray}}
\newcommand{\eea}{\end{eqnarray}}
\newcommand{\be}{\begin{equation}}
\newcommand{\ee}{\end{equation}}

\begin{document}

\begin{frontmatter}

\title{Regular hairy black holes by gravitational decoupling: de Sitter and Minkowski cores}

\author[aff1]{Daulet Berkimbayev\corref{cor1}}
\ead{daulet9432@gmail.com}
\cortext[cor1]{Corresponding author.}
\address[aff1]{Al-Farabi Kazakh National University, 71 Al-Farabi Ave., Almaty 050040, Kazakhstan}

\begin{abstract}
We construct and compare two regular hairy black hole families generated by gravitational decoupling with the same exponentially localized anisotropic sector.
The first branch is built from a Bardeen-type seed and has a de Sitter-like core, whereas the second is generated from an exponential hollow seed and has a Minkowski-like core with vanishing central curvature.
For both branches we determine the static and Kerr-like rotating horizon structure, including the critical deformations that separate two-horizon, extremal, and horizonless geometries.
We prove regularity analytically through curvature-invariant limits, identify the physical parameter domains, and derive the leading asymptotic corrections that distinguish the two cores.
For representative black hole configurations the total effective source satisfies the weak energy condition outside the event horizon.
We also compute the Hawking temperature, entropy, fixed hair heat capacity, photon sphere radius, and static shadow radius, showing that the two regularization mechanisms lead to different thermodynamic and optical trends even when the deformation profile is held fixed.

\end{abstract}


\end{frontmatter}

\section{Introduction}

Curvature singularities are one of the clearest indications that the classical description of black holes is incomplete.
The Penrose--Hawking singularity theorems show that geodesic incompleteness follows under broad geometric and energy assumptions \cite{Penrose:1969pc,Hawking:1973uf}.
Cosmic censorship then separates two related issues: weak cosmic censorship asks whether singularities formed in collapse remain hidden behind event horizons, while strong cosmic censorship concerns the stability of deterministic evolution in the presence of Cauchy horizons \cite{Wald:1997gcc,VandeMoortel:2025scc}.
Regular black hole models provide effective geometries in which the central singularity is removed, usually at the price of introducing nontrivial matter sectors or modified gravitational dynamics.

Many regular geometries have been constructed by coupling gravity to effective nonlinear sources, by modifying the short-distance behavior of the mass function, or by using phenomenological metrics designed to keep curvature invariants finite \cite{Burinskii:2001bq,Dymnikova:2006wn,Smailagic:2010nv,Bambi:2013ufa,Azreg-Ainou:2014nra,Dymnikova:2016nlb,Salazar:1987ap,Ayon-Beato:1998hmi,Bronnikov:2000vy,Balart:2014cga,Toshmatov:2014nya,Fan:2016hvf,Allahyari:2019jqz}.
Energy conditions remain an important diagnostic because they indicate how much exotic effective stress energy is required to support a nonsingular center \cite{Martin-Moruno:2017exc}.
At the same time, black hole hair provides a natural language for parametrizing deviations from the Kerr or Schwarzschild families. Hairy black holes occur in scalar-tensor theories, higher curvature models, and other generalized interactions \cite{Sotiriou:2011dz,Babichev:2013cya,Sotiriou:2013qea,Herdeiro:2014goa,Anabalon:2013eqa,Anabalon:2015xvl,Kleihaus:2015iea,Doneva:2017bvd,Benkel:2016rlz,Martinez:2004nb,Antoniou:2017acq,Antoniou:2017hxj}.

Gravitational decoupling (GD) offers a practical way to construct such configurations while retaining analytic control of the field equations \cite{Ovalle:2017fgl,Ovalle:2019qyi,Ovalle:2018umz,Ovalle:2020kpd,daRocha:2020gee,Ovalle:2021jzf,Ovalle:2022eqb,Ramos:2021jta,Macedo:2021uah,Afrin:2021ggx,Mahapatra:2022xea,Heydarzade:2023dof}.
In the implementation introduced in Ref.~\cite{Ovalle:2023vvu}, a seed geometry is supplemented by an anisotropic tensor-vacuum sector whose profile can be chosen so that the exterior weak energy condition (WEC) is under control.
The same construction also admits a Kerr-like extension through a radially dependent mass function \cite{Contreras:2021yxe,Gurses:1975vu}.

The purpose of this work is to use one fixed decoupling profile to compare two different regular cores.
The first is a Bardeen-type seed, which is already regular and has a de Sitter-like center.
The second is an exponential hollow seed, whose mass function vanishes faster than any power as $r\to0^+$ and therefore gives a Minkowski-like center.
This seed should not be confused with the canonical Simpson--Visser black-bounce geometry; the reference to a hollow core here means only that the central mass profile and central curvature vanish, while black-bounce geometries provide a different well-known route to nonsingular cores \cite{Simpson:2019hgc}.
Keeping the hair profile fixed makes the comparison transparent, that any difference in the phase structure, curvature behavior, thermodynamics, or photon sphere scale can be traced to the underlying seed core.

Our main results are as follows.
First, we determine the static and rotating horizon phase structure of both branches and give the critical deformation values at which double horizons form.
Second, we prove central regularity analytically using curvature-invariant limits and identify the parameter domain in which the fixed-ADM Bardeen branch keeps a positive seed mass.
Third, we show that the Bardeen branch first differs from Schwarzschild at order $r^{-3}$, while the hollow branch contains a Reissner--Nordstr\"om-like $r^{-2}$ asymptotic correction.
Finally, we compute the Hawking temperature, fixed-hair heat capacity, photon sphere radius, and static shadow radius, providing simple physical diagnostics of the two regularization mechanisms.

Recent work on pre-collapse and interior descriptions of regular black holes emphasizes that regularity of a static metric does not by itself solve the dynamical problem \cite{Ovalle:2024BeforeCollapse,Ovalle:2025InteriorDynamics}.
For this reason, throughout the paper we keep the distinction between background regularity, exterior energy conditions, and dynamical stability explicit.
The latter is left for future perturbative analysis.

\section{Gravitational decoupling}
\label{sec:GDnutshell}

For completeness we briefly summarize the GD mechanism used throughout the paper. Detailed discussions can be found in \cite{Ovalle:2017fgl,Ovalle:2019qyi,Ovalle:2023vvu}.
We consider static and spherically symmetric spacetimes,
\begin{equation}
ds^{2}=-e^{\nu(r)}dt^{2}+e^{\lambda(r)}dr^{2}+r^{2}d\Omega^{2},
\label{eq:sss}
\end{equation}
and split the total stress-energy tensor as
\begin{equation}
T_{\mu\nu}=\bar T_{\mu\nu}+\alpha\,\theta_{\mu\nu},
\label{eq:splitGD}
\end{equation}
where $\bar T_{\mu\nu}$ corresponds to a chosen seed configuration and $\theta_{\mu\nu}$ is an additional anisotropic sector that is to be decoupled.
The dimensionless parameter $\alpha$ controls the strength of the deformation. As $\alpha\to0$ we recover the seed.

The Einstein equations then separate into two coupled subsystems.
The first subsystem is the seed Einstein system driven by $\bar T_{\mu\nu}$ and solved by a known seed metric $\{\bar\nu,\bar\lambda\}$.
The second subsystem determines $\theta_{\mu\nu}$ and the deformation of the metric potentials induced by the coupling $\alpha$.
In the minimal geometric deformation (MGD) and related GD schemes, the deformation is chosen so that one of the metric functions is modified in a controlled way, while the remaining field equations become algebraic or reducible to a single quadrature.

In this work we focus on the Kerr--Schild one function subclass employed in \cite{Ovalle:2023vvu}, in which
\begin{equation}
e^{\nu(r)}=e^{-\lambda(r)}\equiv f(r),
\label{eq:onefunction}
\end{equation}
so that the geometry is encoded in a single metric function $f(r)$.
It is convenient to introduce an effective mass function $\tilde m(r)$ defined by
\begin{equation}
f(r)=1-\frac{2\tilde m(r)}{r}.
\label{eq:massdef}
\end{equation}
The GD construction amounts to writing
\begin{equation}
\tilde m(r)=m_{\rm seed}(r)+m_{\rm hair}(r),
\label{eq:mtot}
\end{equation}
where $m_{\rm seed}(r)$ is fixed by the chosen seed and $m_{\rm hair}(r)$ encodes the decoupling sector.
Once a profile for the hair sector density is specified, $m_{\rm hair}(r)$ follows from a single radial integral, and the remaining effective pressures are determined by the Einstein equations.

\section{Static construction and horizon structure}
\label{sec:GD}

For each branch and fixed seed parameters, the horizon structure changes as $\alpha$ is varied.
We define two derived threshold values.
The \textbf{static critical value} $\alpha^*$ is the value for which $f(r)$ develops a degenerate horizon,
\begin{equation}
f(r_e,\alpha^*)=0,\qquad \partial_r f(r_e,\alpha^*)=0 .
\label{eq:static_double_root}
\end{equation}
In terms of the total mass function, these two equations are equivalently
\begin{equation}
\tilde m(r_e)=\frac{r_e}{2},
\qquad
\tilde m'(r_e)=\frac{1}{2}.
\label{eq:static_double_root_mass}
\end{equation}
The \textbf{rotating critical value} $\alpha^{**}$ is the analogous value defined in the rotating extension through $\Delta(r)$ (see Sec.~\ref{sec:rot}),
\begin{equation}
\Delta(r_e,\alpha^{**};a)=0,
\qquad
\partial_r\Delta(r_e,\alpha^{**};a)=0 .
\label{eq:rot_double_root_intro}
\end{equation}
The sign of the phase diagnostic determines the number of horizons, but the ordering of the regimes with respect to $\alpha$ is branch dependent.
For the fixed-ADM Bardeen branch below, increasing $\alpha$ removes the horizons, whereas for the hollow branch with fixed seed mass, increasing $\alpha$ creates them.
Therefore $\alpha$ is the free deformation amplitude, while $\alpha^*$ and $\alpha^{**}$ are critical values determined by the seed parameters and by the chosen rotation parameter.

\begin{figure}[t]
\centering
\includegraphics[width=0.48\linewidth]{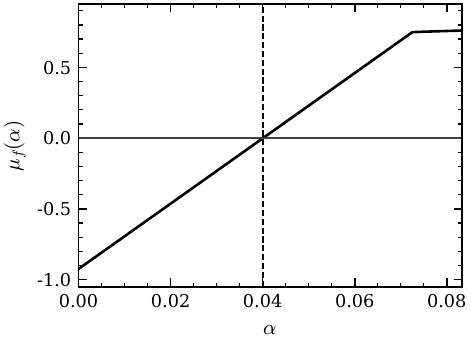}\hfill
\includegraphics[width=0.48\linewidth]{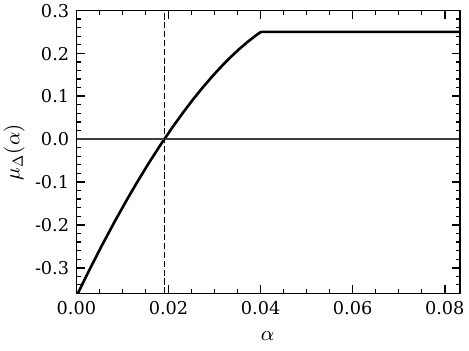}\\[2mm]
\includegraphics[width=0.48\linewidth]{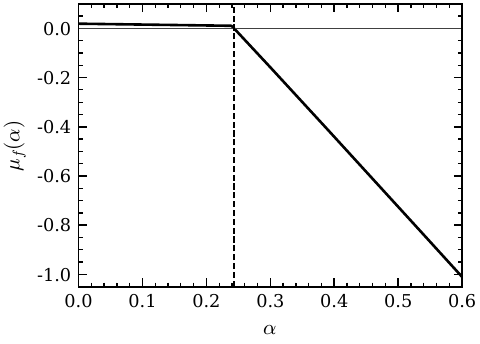}\hfill
\includegraphics[width=0.48\linewidth]{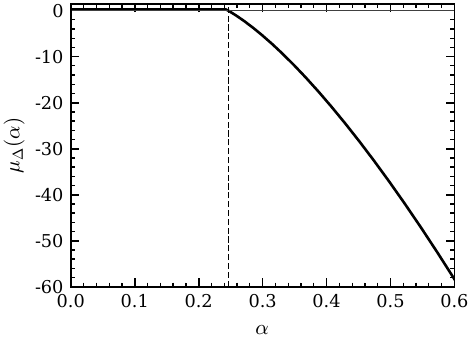}
\caption{Phase-diagram diagnostics for horizon formation.
Top row: Bardeen seed at fixed ADM mass, restricted to the positive seed mass domain $0\le\alpha<M_0/(12\ell)$; bottom row: hollow seed with fixed seed mass.
Left column: $\mu_f(\alpha)=\min_r f(r,\alpha)$ (static); right column: $\mu_\Delta(\alpha)=\min_r\Delta(r,\alpha)$ (rotating, with $a=0.5$).
In all panels, a positive minimum implies no horizon, a zero minimum corresponds to the extremal configuration, and a negative minimum implies two horizons.
The vertical dashed lines mark $\alpha^*$ or $\alpha^{**}$.}
\label{fig:phase}
\end{figure}

Ref.~\cite{Ovalle:2023vvu} adopts the WEC as the main physical requirement for the additional source.
To avoid ambiguity, we distinguish the hair contribution from the total effective source.
In this paper the parameter $\alpha$ is included in the hair sector density itself,
\begin{equation}
\label{eq:Eprofile}
\kappa\rho_{\rm hair}(r)
=
\frac{\alpha\,r^2}{\ell^4}\,e^{-r/\ell}.
\end{equation}
The profile is positive for $\alpha>0$ and decays exponentially at infinity.
For the one function geometry \eqref{eq:onefunction}, the effective source associated with any total mass function $\tilde m(r)$ satisfies
\begin{equation}
\kappa\rho_{\rm eff}=\frac{2\tilde m'}{r^2},
\qquad
\kappa p_{r,{\rm eff}}=-\frac{2\tilde m'}{r^2},
\qquad
\kappa p_{t,{\rm eff}}=-\frac{\tilde m''}{r}.
\label{eq:effective_sources}
\end{equation}
Thus $\rho_{\rm eff}+p_{r,{\rm eff}}=0$ is saturated identically, and the nontrivial WEC inequalities are
\begin{equation}
\rho_{\rm eff}\ge 0,
\qquad
\rho_{\rm eff}+p_{t,{\rm eff}}\ge 0.
\label{eq:WEC}
\end{equation}
The WEC plots in Sec.~\ref{sec:energy_discussion} refer to these total effective quantities.
For comparison, the hair sector alone obeys
\begin{equation}
\kappa(\rho_{\rm hair}+p_{t,{\rm hair}})
=
\frac{\alpha r^2}{2\ell^4}e^{-r/\ell}\left(\frac{r}{\ell}-2\right),
\end{equation}
so the hair sector WEC by itself is guaranteed only for $r\ge 2\ell$.

It is useful to rewrite the deformation in terms of a mass contribution.
In the one function class \eqref{eq:massdef}, the field equation implies
\begin{equation}
\tilde m'(r) = \frac{\kappa}{2}\,\rho_{\rm eff}(r)\,r^2.
\label{eq:mdot}
\end{equation}
For the profile \eqref{eq:Eprofile}, the associated hair mass contribution is
\begin{align}
m_{\rm hair}(r)
&=\frac{1}{2}\int_0^{r} \kappa \rho_{\rm hair}(\rho)\,\rho^2\,d\rho \notag
=\alpha\Big[
12\ell
\\
&-\frac{e^{-r/\ell}}{2\ell^3}\Big(r^4+4\ell r^3+12\ell^2 r^2+24\ell^3 r+24\ell^4\Big)
\Big].
\label{eq:mhair}
\end{align}

For the static configurations we use the phase diagnostic
\begin{equation}
\mu_f(\alpha)\equiv \min_{r>0} f(r,\alpha),
\end{equation}
and for the Kerr-like rotating extension we analogously define
\begin{equation}
\mu_{\Delta}(\alpha)\equiv \min_{r>0}\Delta(r,\alpha).
\end{equation}
In both cases a positive minimum implies no horizons, a zero minimum gives the extremal value, and a negative minimum implies two horizons.
The resulting phase curves for both the Bardeen and hollow branches are shown in Fig.~\ref{fig:phase}.

In all branches we use the same final form
\begin{equation}
f(r) = 1 - \frac{2}{r}\Big(m_{\rm seed}(r)+m_{\rm hair}(r)\Big),
\label{eq:f_general}
\end{equation}
and vary only the seed mass function $m_{\rm seed}(r)$.

For a Schwarzschild seed, $m_{\rm seed}(r)=M$.
This is the starting point of the simplest branch in Ref.~\cite{Ovalle:2023vvu}.
A Schwarzschild seed has no de Sitter core, but it is singular at $r=0$.
This motivates considering regular seeds from the start like Bardeen and also seeds with a Minkowski-like core (H), while keeping the same deformation profile \eqref{eq:Eprofile}.

\subsection{Bardeen seed. Regular solution with de Sitter core}
\label{subsec:bardeen}

A common regular seed is the Bardeen-type mass function
\begin{equation}
m_{\rm seed}(r) \equiv m_{\rm B}(r)
=
M\,\frac{r^3}{\big(r^2+g^2\big)^{3/2}},
\label{eq:bardeen_mass}
\end{equation}
with parameter $g$.
Near the center, $m_{\rm B}(r)\sim (M/g^3)\,r^3$, so the metric behaves as
\begin{equation}
f(r) = 1 - \frac{2M}{g^3}\,r^2 + \mathcal O(r^4),
\label{eq:bardeen_core}
\end{equation}
which corresponds to a de Sitter-like core.

For the profile \eqref{eq:Eprofile} one finds that the hair contribution approaches a constant,
$m_{\rm hair}(r)\to 12\alpha \ell$ as $r\to\infty$.
If the seed mass parameter were kept fixed, the ADM mass of the deformed solution would therefore vary with $\alpha$.
In the Bardeen branch we instead compare configurations at fixed asymptotic mass by renormalizing the seed parameter according to
\begin{equation}
M_{\rm B}(\alpha)\equiv M_{\rm seed}(\alpha)=M_0-12\alpha \ell,
\label{eq:MseedB}
\end{equation}
so that the total mass function satisfies $\tilde m(r)\to M_0$ at infinity for all $\alpha$.
This choice makes the deformation parameter $\alpha$ a hair strength at fixed ADM data.

\begin{figure}[!t]
\centering
\includegraphics[width=0.48\linewidth]{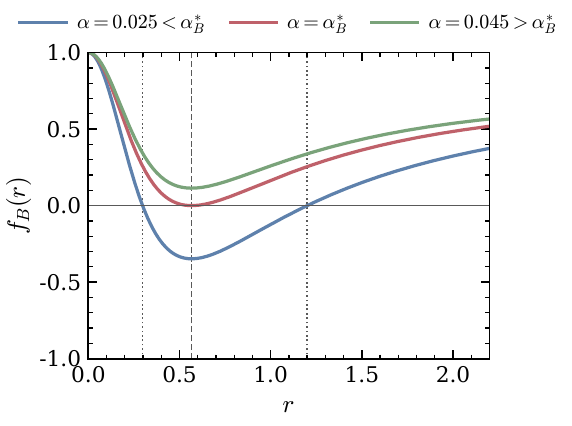}
\hfill
\includegraphics[width=0.48\linewidth]{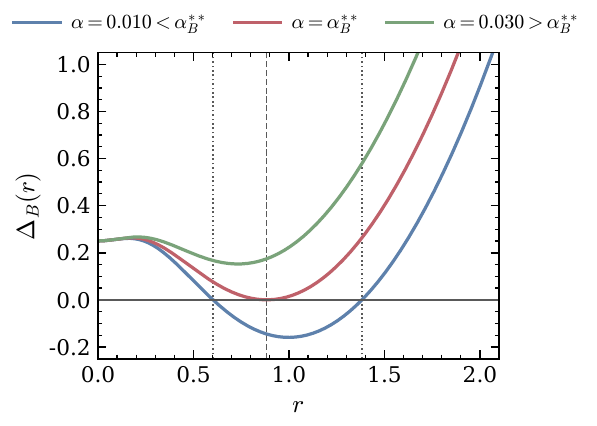}
\caption{Bardeen-seed branch. Left: metric function $f_{\rm B}(r)=e^{-\lambda}=e^{\nu}$ for $\alpha^{\rm B}_1=0.025$ (two horizons), $\alpha^{\rm B}_2=\alpha^{*}_{\rm B}$ (extremal) and $\alpha^{\rm B}_3=0.045$ (horizonless). Right: rotating extension described by $\Delta_{\rm B}(r)$ for $a_{\rm B}=0.5$; the ordering is two horizons, extremal, horizonless as $\alpha$ crosses $\alpha^{**}_{\rm B}$. Vertical dotted/dashed lines mark the corresponding horizon positions for the representative black hole and extremal cases.}
\label{fig:bardeen_metric}
\end{figure}

For the representative choice
\begin{equation}
M_0=1,\qquad g=0.4,\qquad \ell=1,
\end{equation}
the static double-root equations \eqref{eq:static_double_root_mass} give
\begin{equation}
\alpha^{*}_{\rm B}=0.0401,
\qquad
r_{e,{\rm B}}=0.5665.
\end{equation}
Because the seed mass parameter is renormalized as $M_{\rm seed}=M_0-12\alpha\ell$, increasing $\alpha$ weakens the Bardeen seed at fixed ADM mass.
Consequently, the ordering of the static regimes is
\begin{equation}
\begin{aligned}
\alpha<\alpha^{*}_{\rm B}&:\quad \text{two horizons},\\
\alpha=\alpha^{*}_{\rm B}&:\quad \text{extremal},\\
\alpha>\alpha^{*}_{\rm B}&:\quad \text{horizonless}.
\end{aligned}
\end{equation}
To display the three regimes in Fig.~\ref{fig:bardeen_metric}, we choose
\begin{equation}
\alpha^{\rm B}_1=0.025<\alpha^{*}_{\rm B},\qquad
\alpha^{\rm B}_2=\alpha^{*}_{\rm B},\qquad
\alpha^{\rm B}_3=0.045>\alpha^{*}_{\rm B}.
\end{equation}
For $\alpha^{\rm B}_1=0.025$ the solution admits two horizons,
\begin{equation}
r_-^{\rm B}=0.2974,
\qquad
r_+^{\rm B}=1.2.
\end{equation}
The configuration $\alpha^{\rm B}_3=0.045$ has no positive real zero of $f_{\rm B}(r)$.

\begin{figure}[!t]
\centering
\includegraphics[width=0.78\linewidth]{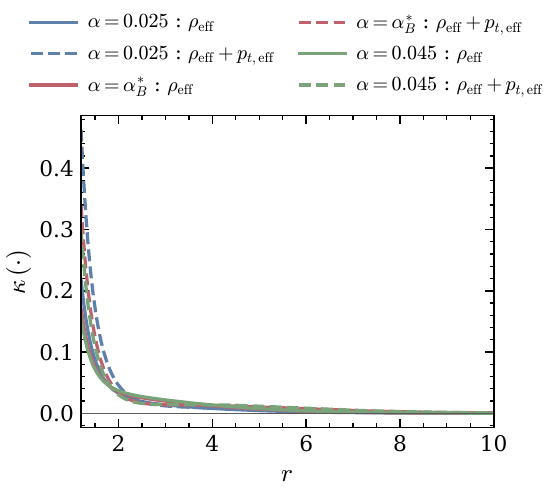}
\caption{Bardeen-seed branch. Total effective WEC combinations for the same deformation strengths used in Fig.~\ref{fig:bardeen_metric}. Solid curves show $\kappa\rho_{\rm eff}$ and dashed curves show $\kappa(\rho_{\rm eff}+p_{t,{\rm eff}})$. The black hole case $\alpha_{\rm B}=0.025$ is plotted outside its outer horizon $r_+^{\rm B}=1.2$; the radial condition $\rho_{\rm eff}+p_{r,{\rm eff}}=0$ is saturated identically.}
\label{fig:bardeen_sources}
\end{figure}

\subsection{Exponential hollow seed with a Minkowski-like core}
\label{subsec:hollow}

To obtain a regular center without a de Sitter core, we use an exponential hollow seed with
\begin{equation}
m_{\rm seed}(r)\equiv m_{\rm H}(r)=M\,e^{-\ell_c/r},
\label{eq:h_mass}
\end{equation}
where $\ell_c$ is a core length scale.
The adjective ``hollow'' is used only in this mass-function sense: the seed mass and all of its derivatives vanish at the center. The metric is not the canonical Simpson--Visser black-bounce line element.
As $r\to 0^+$, $e^{-\ell_c/r}$ decays faster than any power, so $m_{\rm H}(r)=o(r^n)$ for all $n$.
This implies $f(r)\to 1$, and since both the hollow seed and the hair profile vanish at the origin, the total effective central density also vanishes. The core is therefore Minkowski-like.

\begin{figure}[!t]
\centering
\includegraphics[width=0.48\linewidth]{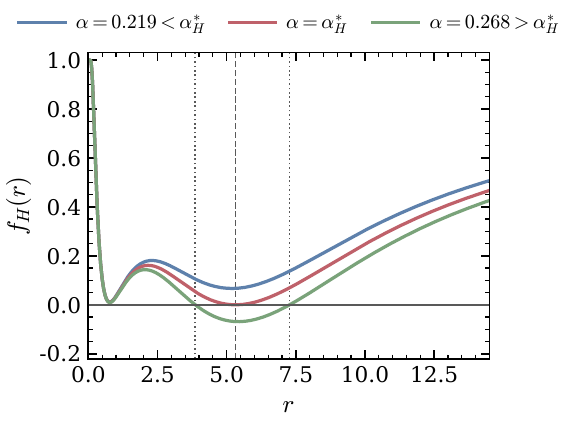}
\hfill
\includegraphics[width=0.48\linewidth]{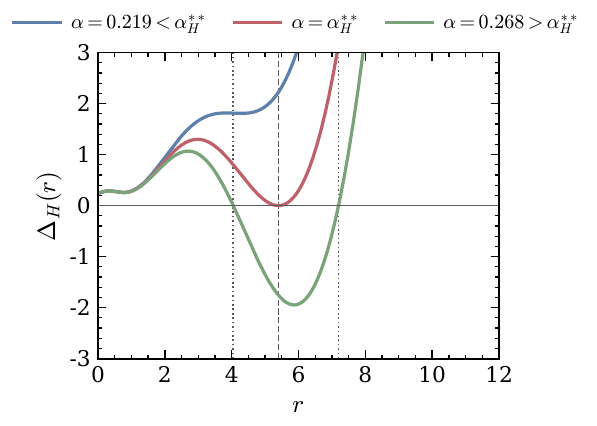}
\caption{Hollow branch. Left: $f_{\rm H}(r)$ for $\alpha^{\rm H}_1=0.219$ (horizonless), $\alpha^{\rm H}_2=\alpha^{*}_{\rm H}$ (extremal) and $\alpha^{\rm H}_3=0.268$ (two horizons). Right: rotating extension described by $\Delta_{\rm H}(r)$ for $a_{\rm H}=0.5$, with the analogous ordering around $\alpha^{**}_{\rm H}$. Vertical dotted/dashed lines mark the corresponding horizon positions for the representative black hole and extremal cases.}
\label{fig:hollow_metric}
\end{figure}

For the representative choice
\begin{equation}
M=1,\qquad \ell_c=0.75,\qquad \ell=1,
\end{equation}
we keep the hollow-seed mass parameter fixed.
Therefore the ADM mass of this branch varies as
\begin{equation}
M_{\rm ADM}^{\rm H}=M+12\alpha\ell.
\label{eq:MADM_H}
\end{equation}
This convention is useful for exhibiting the hollow-branch horizon transition, but it should be kept distinct from the fixed-ADM Bardeen comparison.
The static double-root equations give
\begin{equation}
\alpha^{*}_{\rm H}=0.2432,
\qquad
r_{e,{\rm H}}=5.3174.
\end{equation}
To display the three regimes in Fig.~\ref{fig:hollow_metric}, we choose
\begin{align}
\alpha^{\rm H}_1&=0.219<\alpha^{*}_{\rm H},\qquad
\alpha^{\rm H}_2=\alpha^{*}_{\rm H},\\
\alpha^{\rm H}_3&=0.268>\alpha^{*}_{\rm H}.
\end{align}
Thus $\alpha^{\rm H}_1$, $\alpha^{\rm H}_2$, and $\alpha^{\rm H}_3$ are respectively horizonless, extremal, and two-horizon cases.
For $\alpha^{\rm H}_3=0.268$ the horizon radii are
\begin{equation}
r_-^{\rm H}=3.8649,
\qquad
r_+^{\rm H}=7.2760.
\end{equation}

\begin{figure}[!t]
\centering
\includegraphics[width=0.78\linewidth]{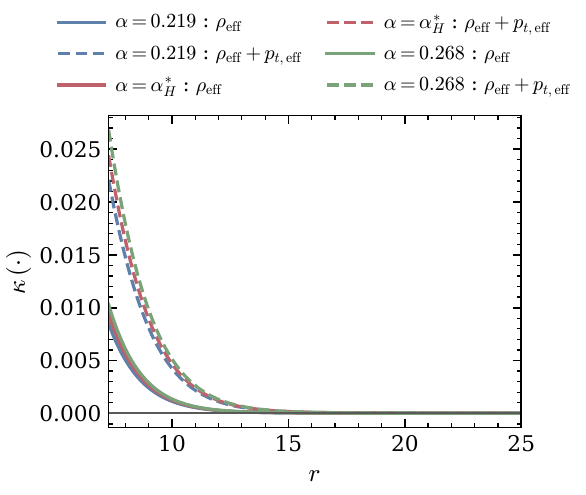}
\caption{Hollow branch. Total effective WEC combinations for the same deformation strengths used in Fig.~\ref{fig:hollow_metric}. Solid curves show $\kappa\rho_{\rm eff}$ and dashed curves show $\kappa(\rho_{\rm eff}+p_{t,{\rm eff}})$. The black hole case $\alpha_{\rm H}=0.268$ is plotted outside its outer horizon $r_+^{\rm H}=7.2760$; the radial WEC combination vanishes identically, $\rho_{\rm eff}+p_{r,{\rm eff}}=0$.}
\label{fig:hollow_sources}
\end{figure}

\section{Regularity and parameter domains}
\label{sec:regularity}

We now record the analytic checks that control central regularity, the parameter domain in which the branches have the intended physical interpretation, and the leading asymptotic corrections. These results make the comparison between the Bardeen and hollow branches independent of the particular numerical plots.

For any static metric in the one function form \eqref{eq:onefunction}--\eqref{eq:massdef}, the scalar curvature invariants can be written directly in terms of the total mass function $\tilde m(r)$:
\begin{align}
R&=\frac{2}{r^2}\left(2\tilde m'(r)+r\tilde m''(r)\right),
\label{eq:R_invariant}\\
R_{\mu\nu}R^{\mu\nu}
&=2\left(\frac{2\tilde m'(r)}{r^2}\right)^2
+2\left(\frac{\tilde m''(r)}{r}\right)^2,
\label{eq:Ricci_square}\\
K&\equiv R_{\mu\nu\rho\sigma}R^{\mu\nu\rho\sigma}
=\left(f''\right)^2+4\left(\frac{f'}{r}\right)^2
+4\left(\frac{1-f}{r^2}\right)^2
\notag\\
&=\frac{4}{r^6}\Big[\left(r^2\tilde m''-2r\tilde m'+2\tilde m\right)^2
\notag\\
&\quad+4\left(\tilde m-r\tilde m'\right)^2+4\tilde m^2\Big].
\label{eq:Kretschmann_mass}
\end{align}
Thus a sufficient static regularity condition is that $\tilde m(r)=O(r^3)$ near $r=0$, with compatible first and second derivatives. The hair contribution is
\begin{equation}
 m_{\rm hair}(r)=\frac{\alpha r^5}{10\ell^4}
 -\frac{\alpha r^6}{12\ell^5}+O(r^7),
\qquad r\to0 .
\label{eq:mhair_smallr}
\end{equation}

For the Bardeen branch, with the fixed-ADM convention \eqref{eq:MseedB}, one obtains
\begin{equation}
\tilde m_{\rm B}(r)=C_{\rm B}r^3+O(r^5),
\qquad
C_{\rm B}=\frac{M_0-12\alpha\ell}{g^3}.
\label{eq:bardeen_C}
\end{equation}
Consequently,
\begin{equation}
f_{\rm B}(r)=1-2C_{\rm B}r^2+O(r^4),
\label{eq:bardeen_f_small}
\end{equation}
and the central invariant limits are finite:
\begin{align}
\lim_{r\to0}R_{\rm B}&=24C_{\rm B},
\label{eq:bardeen_invariant_limits}\\
\lim_{r\to0}\left(R_{\mu\nu}R^{\mu\nu}\right)_{\rm B}
&=144C_{\rm B}^2,\notag\\
\lim_{r\to0}K_{\rm B}&=96C_{\rm B}^2.\notag
\end{align}
For $C_{\rm B}>0$ the core is de Sitter-like, with effective cosmological constant $\Lambda_{\rm eff}=6C_{\rm B}$. At $C_{\rm B}=0$ the Bardeen seed no longer supplies the leading core term and the hair sector makes the center Minkowski-like at leading order; for $C_{\rm B}<0$ the seed is a formal negative mass, anti-de Sitter-like core contribution.

For the hollow branch, $m_{\rm H}(r)=M\exp(-\ell_c/r)$ and all of its derivatives vanish faster than any power as $r\to0^+$. Combining this with \eqref{eq:mhair_smallr} gives
\begin{equation}
\tilde m_{\rm H}(r)=\frac{\alpha r^5}{10\ell^4}+O(r^6)+o(r^N)
\quad \text{for every fixed }N,
\label{eq:h_mass_small}
\end{equation}
so that
\begin{equation}
f_{\rm H}(r)=1-\frac{\alpha r^4}{5\ell^4}+O(r^5)+o(r^N),
\qquad r\to0^+.
\label{eq:h_f_small}
\end{equation}
Equations \eqref{eq:R_invariant}--\eqref{eq:Kretschmann_mass} then imply
\begin{align}
\lim_{r\to0}R_{\rm H}&=0,\notag\\
\lim_{r\to0}\left(R_{\mu\nu}R^{\mu\nu}\right)_{\rm H}&=0,
\label{eq:h_invariant_limits}\\
\lim_{r\to0}K_{\rm H}&=0.
\notag
\end{align}
The hollow branch therefore has an asymptotically Minkowski core rather than a de Sitter core.

\begin{figure}[t]
\centering
\includegraphics[width=0.92\linewidth]{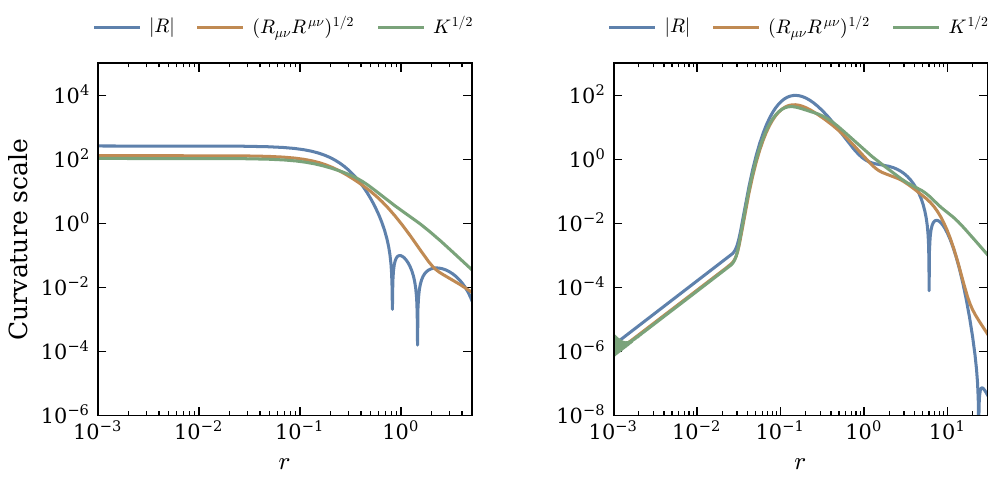}
\caption{Curvature scales for representative two-horizon geometries. The left panel uses the Bardeen branch with $\alpha_{\rm B}=0.025$ and the right panel uses the hollow branch with $\alpha_{\rm H}=0.268$. We plot $|R|$, $\sqrt{R_{\mu\nu}R^{\mu\nu}}$, and $\sqrt{K}$ to show that the invariants remain finite near the core. The Bardeen curves approach finite nonzero central values, whereas the hollow-branch curves tend to zero, as predicted by Eqs.~\eqref{eq:bardeen_invariant_limits} and \eqref{eq:h_invariant_limits}.}
\label{fig:curvature_scales}
\end{figure}

\subsection{Physical parameter domains}
\label{subsec:param_domains}

The hair sector density \eqref{eq:Eprofile} is non-negative for $\alpha\ge0$ and $\ell>0$. The seed parameters are taken as $g>0$ for the Bardeen branch and $\ell_c>0$ for the hollow branch. The additional restriction depends on the mass convention used in each branch.

\begin{table}[t]
\caption{Parameter domains used in the analytic discussion. The Bardeen branch is compared at fixed ADM mass, while the hollow branch in the present numerical examples keeps the seed mass $M$ fixed and therefore has $M_{\rm ADM}^{\rm H}=M+12\alpha\ell$.}
\label{tab:parameter_domains}
\centering
\small
\renewcommand{\arraystretch}{1.18}
\begin{tabularx}{\linewidth}{>{\raggedright\arraybackslash}p{0.22\linewidth}>{\raggedright\arraybackslash}p{0.21\linewidth}>{\centering\arraybackslash}p{0.18\linewidth}>{\raggedright\arraybackslash}X}
\toprule
Branch/convention & Basic assumptions & Domain for $\alpha$ & Interpretation \\
\midrule
Bardeen, fixed ADM & $M_0>0$, $g>0$, $\ell>0$ & $0\le\alpha<M_0/(12\ell)$ & Positive seed mass and de Sitter-like core \\
Bardeen, boundary value & As above & $\alpha=M_0/(12\ell)$ & Seed core coefficient $C_{\rm B}=0$ \\
Bardeen, formal extension & As above & $\alpha>M_0/(12\ell)$ & Negative seed mass; formal branch \\
Hollow, fixed seed mass & $M>0$, $\ell_c>0$, $\ell>0$ & $\alpha\ge0$ & Minkowski core, $M_{\rm ADM}=M+12\alpha\ell$ \\
Hollow, optional fixed ADM comparison & $M_0>0$, $\ell_c>0$, $\ell>0$ & $0\le\alpha<M_0/(12\ell)$ & Obtained by setting $M=M_0-12\alpha\ell$ \\
\bottomrule
\end{tabularx}
\end{table}

For the numerical Bardeen choice $M_0=1$ and $\ell=1$, the positive seed mass upper bound is
\begin{equation}
\alpha_{\rm max}^{\rm B}=\frac{M_0}{12\ell}=0.0833\ldots .
\end{equation}
The representative Bardeen values in Table~\ref{tab:horizon_data}, including $\alpha_{\rm B}^*=0.0401$ and $\alpha_{\rm B}^{**}=0.0191$, lie inside this physical domain. This is why the Bardeen phase plots in Fig.~\ref{fig:phase} are restricted to the positive-seed-mass range.

\subsection{Asymptotic structure}
\label{subsec:asymptotics}

At large radius the hair mass approaches
\begin{equation}
m_{\rm hair}(r)=12\alpha\ell+O\!\left(r^4e^{-r/\ell}\right),
\qquad r\to\infty .
\label{eq:mhair_large}
\end{equation}
Thus the Bardeen fixed-ADM branch has $M_{\rm ADM}^{\rm B}=M_0$. Its large-$r$ expansion is
\begin{align}
\tilde m_{\rm B}(r)
&=M_0-\frac{3M_{\rm B}(\alpha)g^2}{2r^2}
+\frac{15M_{\rm B}(\alpha)g^4}{8r^4}+O(r^{-6})
\notag\\
&\quad+O\!\left(r^4e^{-r/\ell}\right),
\label{eq:bardeen_asymp_m}\\
f_{\rm B}(r)
&=1-\frac{2M_0}{r}
+\frac{3M_{\rm B}(\alpha)g^2}{r^3}
-\frac{15M_{\rm B}(\alpha)g^4}{4r^5}
\notag\\
&\quad+O(r^{-7})+O\!\left(r^3e^{-r/\ell}\right).
\label{eq:bardeen_asymp_f}
\end{align}
Therefore the first algebraic correction beyond Schwarzschild appears at order $r^{-3}$.

For the hollow branch with fixed seed mass,
\begin{equation}
M_{\rm ADM}^{\rm H}=M+12\alpha\ell,
\end{equation}
and
\begin{align}
\tilde m_{\rm H}(r)
&=M_{\rm ADM}^{\rm H}-\frac{M\ell_c}{r}
+\frac{M\ell_c^2}{2r^2}
\notag\\
&\quad-\frac{M\ell_c^3}{6r^3}+O(r^{-4})
+O\!\left(r^4e^{-r/\ell}\right),
\label{eq:h_asymp_m}\\
f_{\rm H}(r)
&=1-\frac{2M_{\rm ADM}^{\rm H}}{r}
+\frac{2M\ell_c}{r^2}
\notag\\
&\quad-\frac{M\ell_c^2}{r^3}
+\frac{M\ell_c^3}{3r^4}+O(r^{-5})
\notag\\
&\quad+O\!\left(r^3e^{-r/\ell}\right).
\label{eq:h_asymp_f}
\end{align}
The hollow branch therefore contains a Reissner--Nordstr\"om-like $r^{-2}$ correction with effective strength
\begin{equation}
Q_{\rm eff}^2=2M\ell_c,
\end{equation}
whereas the Bardeen branch does not. This difference provides a clean asymptotic way to distinguish the two regularization mechanisms.

\subsection{Power-counting check for the Kerr-like extension}
\label{subsec:rot_regularity}

In a Kerr-like mass function geometry, the possible Kerr-type ring divergence is associated with the region $\Sigma=r^2+a^2\cos^2\theta\to0$. We therefore use power counting near this region as a necessary local regularity check. In the present mass-function construction the potentially dangerous terms are controlled by $r\tilde m(r)/\Sigma$ and by derivatives of $\tilde m(r)$ divided by powers of $\Sigma$. A sufficient condition for the absence of the Kerr $1/\Sigma$-type divergence, as $r\to0$, is
\begin{equation}
\tilde m(r)=O(r^3),\qquad
\tilde m'(r)=O(r^2),\qquad
\tilde m''(r)=O(r).
\label{eq:rot_reg_condition}
\end{equation}
The Bardeen branch satisfies this condition through \eqref{eq:bardeen_C}, while the hollow branch is even softer, with \eqref{eq:h_mass_small}. This indicates that the Kerr-like replacement does not reintroduce the usual ring-type power divergence at the level of curvature power counting. This regularity statement is separate from dynamical stability, which requires a perturbative analysis. The thermodynamic and photon sphere observables of the static branches are analysed below, while a full dynamical stability analysis remains beyond the scope of the present work.

\section{Rotating extension}
\label{sec:rot}

Following Ref.~\cite{Contreras:2021yxe} and the Kerr--Schild strategy
\cite{Gurses:1975vu}, we construct a Kerr-like rotating extension by
promoting the Kerr mass parameter $M$ to a radially deformed mass
function $\tilde m(r)$. In Boyer--Lindquist--type coordinates
$(t,r,\theta,\phi)$ we introduce the standard metric functions
\begin{align}
\Sigma(r,\theta) &= r^2 + a^2\cos^2\theta,\\
\Delta(r) &= r^2 - 2 r\,\tilde m(r) + a^2 .
\end{align}
The corresponding rotating line element can be written in the Kerr form
\begin{align}
ds^2 &= -\left(1-\frac{2 r\,\tilde m(r)}{\Sigma}\right)dt^2
-\frac{4 a r\,\tilde m(r)\sin^2\theta}{\Sigma}\, dt\, d\phi\\
&\quad+\frac{\Sigma}{\Delta}\,dr^2 + \Sigma\, d\theta^2 \nonumber\\
&\quad + \sin^2\theta\left(r^2+a^2+\frac{2 a^2 r\,\tilde   m(r)\sin^2\theta}{\Sigma}\right)d\phi^2 \nonumber.
\end{align}

In our setup the deformation is branch-dependent. For each branch
$i\in\{{\rm B},{\rm H}\}$ we take
\begin{equation}
\Delta_i(r)=r^{2}-2r\,\tilde m_i(r)+a_i^{2},
\label{eq:Delta}
\end{equation}
where $\tilde m_i(r)=m_{{\rm seed},i}(r)+m_{{\rm hair},i}(r)$ and the
rotation parameter $a_i$ may be chosen independently for the two
branches. The rotating horizons are determined by the real positive
roots of $\Delta_i(r)=0$, while extremality occurs when
$\Delta_i(r_h)=0$ and $\Delta_i'(r_h)=0$.

At the critical value $\alpha^{**}_i$ the root becomes degenerate, so that $\Delta_i(r_e)=0$ and $\Delta_i'(r_e)=0$.
Equivalently,
\begin{equation}
\tilde m_i(r_e)+r_e\tilde m_i'(r_e)=r_e,
\qquad
 a_i^2=2r_e\tilde m_i(r_e)-r_e^2.
\label{eq:rot_ext_mass}
\end{equation}
For the Bardeen branch we use
\begin{equation}
a_{\rm B}=0.5,
\qquad
\alpha^{**}_{\rm B}=0.0191,
\qquad
r^{\rm rot}_{e,{\rm B}}=0.8828.
\end{equation}
For this fixed-ADM branch, $\alpha<\alpha^{**}_{\rm B}$ gives two rotating horizons, whereas $\alpha>\alpha^{**}_{\rm B}$ gives no horizon.
For the hollow branch we use
\begin{equation}
a_{\rm H}=0.5,
\qquad
\alpha^{**}_{\rm H}=0.2464,
\qquad
r^{\rm rot}_{e,{\rm H}}=5.3983.
\end{equation}
For the hollow branch the ordering is the standard one with respect to $\alpha$: $\alpha<\alpha^{**}_{\rm H}$ is horizonless and $\alpha>\alpha^{**}_{\rm H}$ gives two rotating horizons.
If one chooses $a_{\rm B}=a_{\rm H}$, the comparison isolates the effect of the seed on the horizon structure at fixed spin.

\begin{table}[t]
\caption{Representative numerical horizon data for the parameter choices used in the figures. Static values use $M_0=1$, $g=0.4$, $\ell=1$ for the Bardeen branch and $M=1$, $\ell_c=0.75$, $\ell=1$ for the hollow branch. Rotating values use $a=0.5$. The symbol $r_e$ denotes a degenerate horizon.}
\label{tab:horizon_data}
\centering
\begin{tabular}{lccccc}
\toprule
Branch & Case & $\alpha$ & $r_-$ & $r_+$ or $r_e$ & Regime \\
\midrule
Bardeen & static   & $0.0250$ & $0.2974$ & $1.2$ & two horizons \\
Bardeen & static   & $0.0401$ & --       & $0.5665$ & extremal \\
Bardeen & static   & $0.0450$ & --       & --       & horizonless \\
Bardeen & rotating & $0.0100$ & $0.6022$ & $1.3825$ & two horizons \\
Bardeen & rotating & $0.0191$ & --       & $0.8828$ & extremal \\
Bardeen & rotating & $0.0300$ & --       & --       & horizonless \\
Hollow  & static   & $0.2190$ & --       & --       & horizonless \\
Hollow  & static   & $0.2432$ & --       & $5.3174$ & extremal \\
Hollow  & static   & $0.2680$ & $3.8649$ & $7.2760$ & two horizons \\
Hollow  & rotating & $0.2190$ & --       & --       & horizonless \\
Hollow  & rotating & $0.2464$ & --       & $5.3983$ & extremal \\
Hollow  & rotating & $0.2680$ & $4.0491$ & $7.1980$ & two horizons \\
\bottomrule
\end{tabular}
\end{table}

\section{Energy conditions}
\label{sec:energy_discussion}

As stated above, the WEC check in this section is applied to the total effective source generated by the full mass function $\tilde m(r)$.
Using \eqref{eq:effective_sources}, the radial combination is saturated,
\begin{equation}
\rho_{\rm eff}+p_{r,{\rm eff}}=0,
\end{equation}
so the nontrivial exterior requirements are
\begin{equation}
\rho_{\rm eff}(r)\ge 0,
\qquad
\rho_{\rm eff}(r)+p_{t,{\rm eff}}(r)\ge 0
\qquad (r\ge r_+).
\end{equation}
Figure~\ref{fig:bardeen_sources} displays the total effective WEC combinations for the three Bardeen deformation strengths used in Fig.~\ref{fig:bardeen_metric}. For the representative two-horizon configuration $\alpha_{\rm B}=0.025$, whose outer horizon is
\begin{equation}
r^{\rm B}_+=1.2,
\end{equation}
the exterior scan over $r\in[r_+^{\rm B},30]$ gives
\begin{align}
\min \kappa\rho_{\rm eff}&=2.7644\times 10^{-8},\\
\min \kappa(\rho_{\rm eff}+p_{t,{\rm eff}})&=6.9122\times 10^{-8}.
\end{align}
Figure~\ref{fig:hollow_sources} similarly displays the total effective WEC combinations for the three hollow-branch deformation strengths used in Fig.~\ref{fig:hollow_metric}. For the representative two-horizon configuration $\alpha_{\rm H}=0.268$, whose outer horizon is
\begin{equation}
r^{\rm H}_+=7.2760,
\end{equation}
the exterior scan over $r\in[r_+^{\rm H},30]$ gives
\begin{align}
\min \kappa\rho_{\rm eff}&=1.8062\times 10^{-6},\\
\min \kappa(\rho_{\rm eff}+p_{t,{\rm eff}})&=3.5900\times 10^{-6}.
\end{align}
In both branches the two plotted WEC combinations remain non-negative outside the event horizon and approach $0^+$ asymptotically.
Thus, on the full exterior domain the quantity has no positive global minimum:
it approaches zero only asymptotically as $r\to\infty$.
The core structure nevertheless differs: de Sitter-like for the Bardeen branch and Minkowski-like for the hollow branch.

\section{Static thermodynamics }
\label{sec:thermo_photon}

We now add two physical diagnostics that are useful for comparing regular black hole geometries: horizon thermodynamics and the photon sphere scale. In this section we restrict attention to the static metrics; the rotating case requires a separate analysis of the full null geodesic system.

\subsection{Horizon thermodynamics}
\label{subsec:thermodynamics}

For a static horizon at $r=r_+$ one has $f(r_+)=0$, or equivalently $\tilde m(r_+)=r_+/2$. The surface gravity gives
\begin{equation}
T_H=\frac{f'(r_+)}{4\pi}
=\frac{1-2\tilde m'(r_+)}{4\pi r_+} .
\label{eq:TH_static}
\end{equation}
Since the geometry is treated as an Einstein-gravity solution supported by an effective source, the entropy is taken to be the Bekenstein--Hawking area entropy,
\begin{equation}
S=\frac{A_+}{4}=\pi r_+^2 .
\label{eq:S_static}
\end{equation}
If the hair and core parameters are held fixed, the ADM mass may be regarded as a function of the horizon radius. Writing
\begin{equation}
m_{\rm hair}(r)=\alpha H_\ell(r),
\end{equation}
where $H_\ell(r)$ is the bracketed function in Eq.~\eqref{eq:mhair}, and defining
\begin{equation}
{\cal B}_g(r)=\frac{r^3}{(r^2+g^2)^{3/2}},
\qquad
{\cal S}_{\ell_c}(r)=e^{-\ell_c/r},
\end{equation}
the horizon equation gives the following mass-radius relations:
\begin{align}
M_{\rm ADM}^{\rm B}(r_+;\alpha)
&=12\alpha\ell+
\frac{r_+/2-\alpha H_\ell(r_+)}{{\cal B}_g(r_+)},
\label{eq:MADM_B_thermo}\\
M_{\rm ADM}^{\rm H}(r_+;\alpha)
&=12\alpha\ell+
\frac{r_+/2-\alpha H_\ell(r_+)}{{\cal S}_{\ell_c}(r_+)} .
\label{eq:MADM_H_thermo}
\end{align}
For the hollow branch this expression makes explicit that the thermodynamic energy is the total ADM mass, even though the phase plots above used a fixed seed mass $M$.
Thus the thermodynamic curve is not a curve at fixed seed mass; rather, it is the family obtained by varying the ADM mass while holding the hair and core parameters fixed.

\begin{figure}[t]
\centering
\includegraphics[width=0.92\linewidth]{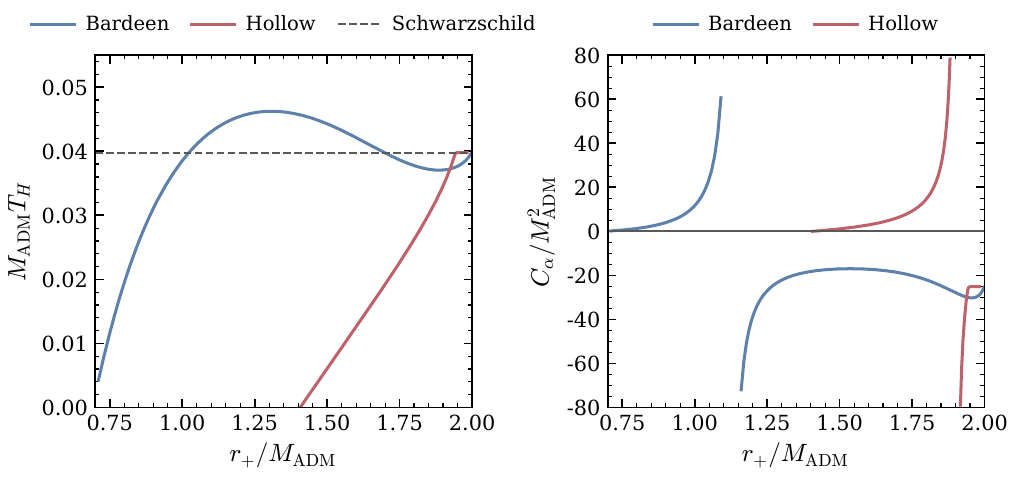}
\caption{Static thermodynamic diagnostics at fixed hair parameters. We use $g=0.4$, $\ell=1$, $\alpha_{\rm B}=0.025$ for the Bardeen branch and $\ell_c=0.75$, $\ell=1$, $\alpha_{\rm H}=0.268$ for the hollow branch. Left: dimensionless Hawking temperature $M_{\rm ADM}T_H$ as a function of $r_+/M_{\rm ADM}$, with the Schwarzschild value $1/(8\pi)$ shown as a dashed line. Right: heat capacity $C_\alpha/M_{\rm ADM}^2$; very large values near heat-capacity poles are omitted from the plotted range to make the sign of the stable and unstable segments visible.}
\label{fig:thermo_static}
\end{figure}

At fixed hair parameters the heat capacity is
\begin{equation}
C_\alpha=\left(\frac{\partial M_{\rm ADM}}{\partial T_H}\right)_{\alpha,g,\ell,\ell_c}
=\frac{dM_{\rm ADM}/dr_+}{dT_H/dr_+} .
\label{eq:C_alpha}
\end{equation}
Poles of $C_\alpha$ occur at extrema of $T_H(r_+)$ and indicate changes between locally thermally stable and unstable branches. Figure~\ref{fig:thermo_static} displays the dimensionless temperature and heat capacity for the representative hair values used in the two-horizon examples. Both branches have a zero-temperature extremal endpoint. The Bardeen branch shown there crosses from positive to negative heat capacity as the horizon grows, whereas the representative hollow-branch black hole in Table~\ref{tab:thermo_photon_values} lies on a positive heat capacity segment.

\subsection{Photon spheres and shadow radius}
\label{subsec:photon_shadow}

For null geodesics in the static metric, the conserved energy and angular momentum lead to
\begin{equation}
\dot r^2+V_{\rm eff}(r)=E^2,
\qquad
V_{\rm eff}(r)=\frac{L^2 f(r)}{r^2} .
\end{equation}
An unstable circular photon orbit satisfies
\begin{equation}
\frac{d}{dr}\left(\frac{f}{r^2}\right)_{r=r_{\rm ph}}=0,
\end{equation}
or
\begin{equation}
r_{\rm ph}f'(r_{\rm ph})-2f(r_{\rm ph})=0 .
\label{eq:photon_condition_f}
\end{equation}
In terms of the mass function this becomes
\begin{equation}
3\tilde m(r_{\rm ph})-r_{\rm ph}\tilde m'(r_{\rm ph})=r_{\rm ph} .
\label{eq:photon_condition_m}
\end{equation}
When several positive roots of Eq.~\eqref{eq:photon_condition_m} exist, we use the outermost root satisfying $r_{\rm ph}>r_+$, since this is the unstable photon orbit relevant for the shadow of an asymptotic observer. Inner roots, when present, lie inside the black hole region or between the horizons and are not used in the optical radius reported below.
For an observer at infinity, the corresponding static shadow radius is the critical impact parameter
\begin{equation}
b_{\rm ph}=\frac{L}{E}=\frac{r_{\rm ph}}{\sqrt{f(r_{\rm ph})}} .
\label{eq:bph}
\end{equation}

\begin{figure}[t]
\centering
\includegraphics[width=0.92\linewidth]{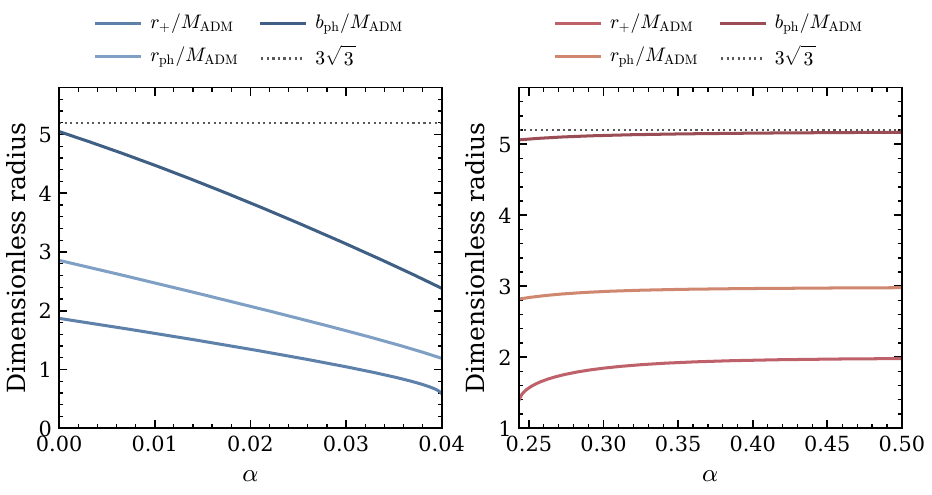}
\caption{Static optical observables. The curves show the event-horizon radius $r_+$, photon sphere radius $r_{\rm ph}$, and shadow radius $b_{\rm ph}$, all normalized by $M_{\rm ADM}$. Left: Bardeen branch at fixed ADM mass $M_{\rm ADM}=M_0=1$ over the black hole interval $0\le\alpha<\alpha_{\rm B}^*$. Right: hollow branch with fixed seed mass $M=1$, so that $M_{\rm ADM}=1+12\alpha\ell$, over the two-horizon interval $\alpha>\alpha_{\rm H}^*$. The dotted line marks the Schwarzschild shadow value $b_{\rm ph}/M_{\rm ADM}=3\sqrt{3}$.}
\label{fig:photon_shadow}
\end{figure}

The resulting trends are shown in Fig.~\ref{fig:photon_shadow}. In the fixed-ADM Bardeen branch, increasing $\alpha$ toward extremality decreases $r_+$, $r_{\rm ph}$, and $b_{\rm ph}$ in ADM units. In the hollow branch, where the representative phase family has $M_{\rm ADM}^{\rm H}=M+12\alpha\ell$, the absolute radii grow with $\alpha$, but the dimensionless ratios remain close to the Schwarzschild scale once normalized by $M_{\rm ADM}^{\rm H}$. This gives a clear observable distinction between the two ways of implementing a regular core with the same deformation profile.

\begin{table}[t]
\caption{Thermodynamic and optical quantities for the representative two-horizon configurations. For Bardeen we use $M_0=1$, $g=0.4$, $\ell=1$, $\alpha=0.025$. For the hollow branch we use $M=1$, $\ell_c=0.75$, $\ell=1$, $\alpha=0.268$, hence $M_{\rm ADM}=4.216$. The heat capacity is evaluated at fixed hair and core parameters using Eq.~\eqref{eq:C_alpha}.}
\label{tab:thermo_photon_values}
\centering
\footnotesize
\renewcommand{\arraystretch}{1.15}
\resizebox{\linewidth}{!}{%
\begin{tabular}{lcccccccc}
\toprule
Branch & $\alpha$ & $M_{\rm ADM}$ & $r_+$ & $T_H$ & $S/M_{\rm ADM}^2$ & $C_\alpha/M_{\rm ADM}^2$ & $r_{\rm ph}/M_{\rm ADM}$ & $b_{\rm ph}/M_{\rm ADM}$ \\
\midrule
Bardeen & $0.025$ & $1.000$ & $1.2000$ & $4.5462\times10^{-2}$ & $4.5238$ & $-38.6786$ & $1.8716$ & $3.4933$ \\
Hollow & $0.268$ & $4.216$ & $7.2760$ & $4.9732\times10^{-3}$ & $9.3570$ & $7.5498$ & $2.8838$ & $5.0986$ \\
\bottomrule
\end{tabular}%
}
\end{table}

\section{Conclusions}

We have constructed two regular hairy black hole branches by applying the same gravitational decoupling profile to two different seed cores.
The Bardeen branch has a de Sitter-like center, while the exponential hollow branch has a Minkowski-like center with vanishing central curvature.
This difference is not only qualitative: the curvature-invariant limits, the allowed physical parameter domains, and the large-radius expansions distinguish the two branches in a controlled way.

The fixed ADM Bardeen branch displays a reversed ordering of the static phase structure: increasing the hair strength weakens the seed mass and removes the horizons. By contrast, in the fixed seed mass hollow branch, increasing the hair strength creates the two-horizon geometry.
The corresponding Kerr-like extensions follow from the same replacement $M\to\tilde m(r)$ and are governed by the double-root conditions of $\Delta(r)$.
For the representative configurations considered here, the total effective source satisfies the WEC outside the event horizon, while the hair sector alone obeys the tangential WEC only for $r\ge2\ell$.

The thermodynamic and optical diagnostics provide an additional physical separation between the two core mechanisms.
Both branches possess zero-temperature extremal endpoints, but their fixed hair heat capacities and shadow scales behave differently.
In the fixed ADM Bardeen branch the photon sphere and shadow radii decrease as the deformation approaches extremality.
In the hollow branch the absolute radii grow with the ADM mass, while the normalized shadow scale remains close to the Schwarzschild value over the representative interval.
These results show that gravitational decoupling can generate regular hairy geometries whose observable trends depend sensitively on the seed core even when the additional anisotropic sector is kept fixed.

The present analysis is limited to background regularity, exterior energy conditions, static thermodynamics, and static photon sphere observables.
A full perturbative stability study, the nonlinear behavior of the inner horizon, and a complete rotating-shadow calculation are natural next steps.

\section*{Data availability}
 No observational data were used in this work. The numerical values and figures are generated from the analytic expressions displayed in the paper.

\bibliographystyle{elsarticle-num}
\bibliography{Hairy_Black_Hole_2}

\end{document}